\documentclass[letterpaper,prl,floatfix,twocolumn,aps,10pt]{revtex4-1}
\usepackage{
    graphicx,
    amsmath
}

\newcommand{\cond}{|}
\newcommand{\bet}{\! : \!}
\pdfoutput=1

\begin{document}


%
\title{On the security of key distribution based on Johnson-Nyquist noise}
\date{\today}
\author{Charles~H.~Bennett, C.~Jess~Riedel}
\affiliation{IBM Watson Research Center, Yorktown Heights, NY, USA}\


\begin{abstract}
We point out that arguments for the security of Kish's noise-based cryptographic protocol have relied on an unphysical no-wave limit, which if taken seriously would prevent any correlation from developing between the users. We introduce a noiseless version of the protocol, also having illusory security in the no-wave limit, to show that noise and thermodynamics play no essential role. Then we prove generally that classical electromagnetic protocols cannot establish a secret key between two parties separated by a spacetime region perfectly monitored by an eavesdropper. We note that the original protocol of Kish is vulnerable to passive time-correlation attacks even in the quasi-static limit. Finally we show that protocols of this type can be secure in practice against an eavesdropper with noisy monitoring equipment. In this case the security is a straightforward consequence of Maurer and Wolf's discovery that key can be distilled by public discussion from correlated random variables in a wide range of situations where the eavesdropper's noise is at least partly independent from the users' noise.
\end{abstract}


\maketitle

\begin{figure} [b!]
    \centering
  \newcommand{\awidthfactor}{0.95}
    \includegraphics[width=\awidthfactor\columnwidth]{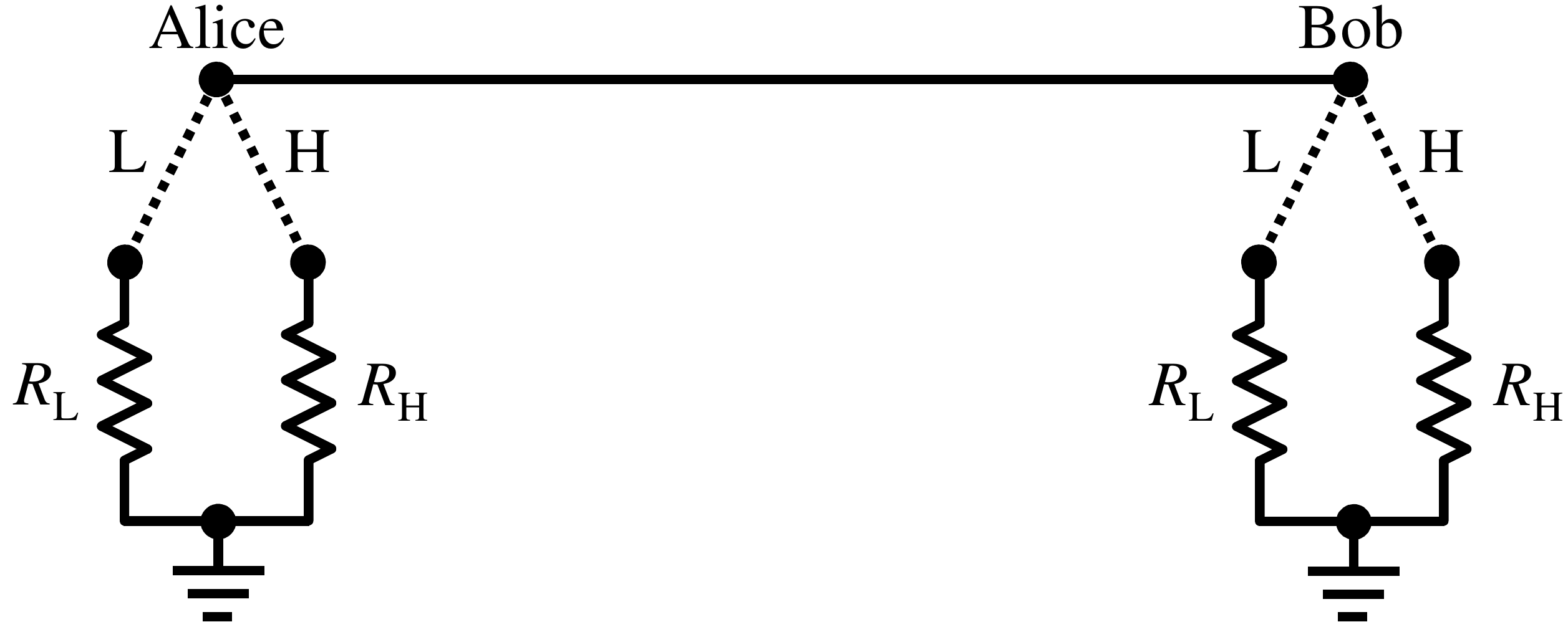}
\caption{\textbf{Basic KLJN protocol.} Alice and Bob are at opposite ends of an ideal transmission line, with Eve somewhere in the middle. At each clock cycle Alice and Bob decide independently and randomly whether to terminate their end of the line with a low resistance (L) or a high resistance (H). The Johnson-Nyquist noise from these thermal resistances $R_\mathrm{L}$ and $R_\mathrm{H}$ at temperature $T$ can be modeled as an ideal resistor in series with a voltage source undergoing Gaussian fluctuations. After transients resulting from this switching have died down, the mean square noise voltage will be the same at any point between Alice and Bob, having one of three values: low (for the LL combination), high (for the HH combination) and intermediate (for LH or HL). Alice and Bob discard data from the HH and LL events, and keep the HL and LH data. The kept events are candidates for a secret key because, knowing their own switch settings, they can distinguish HL from LH, but Eve cannot do so from the mean square noise voltage on the transmission line.}
\label{fig:originalkish}
\end{figure}

Quantum key distribution \cite{Bennett1984, *Shor2000, *Gisin2002} boasts unconditional security even in the presence of realistic noise \cite{Mayers2001, *Gottesman2004, *Lo2012, *Braunstein2012}, and the techniques have matured enough that small commercial implementations have been explored. However, the practical difficulty of manipulating individual quantum states has prompted some investigation into purely classical schemes which might be able to achieve similar ends. In particular, Kish has proposed a strictly classical protocol (Kirchoff Law-Johnson Noise or KLJN) on an insecure transmission line using Johnson-Nyquist noise in resistors \cite{Kish2006a} (figure \ref{fig:originalkish}). He and collaborators claim that the idealized model of this protocol is unconditionally secure and that the prospects of real-world implementations are promising \cite{Kish2011, *Kish2012}. Intuitively, this seems dubious; information passing between two parties must travel as an electromagnetic wave over the transmission line, and a potential eavesdropper should be able to perfectly measure a classical signal (regardless of its noisy origins) with arbitrary precision and with arbitrarily little disturbance.

\begin{figure} [b!]
\centering
\newcommand{\widthfactor}{0.95}
\includegraphics[width=\widthfactor\columnwidth]{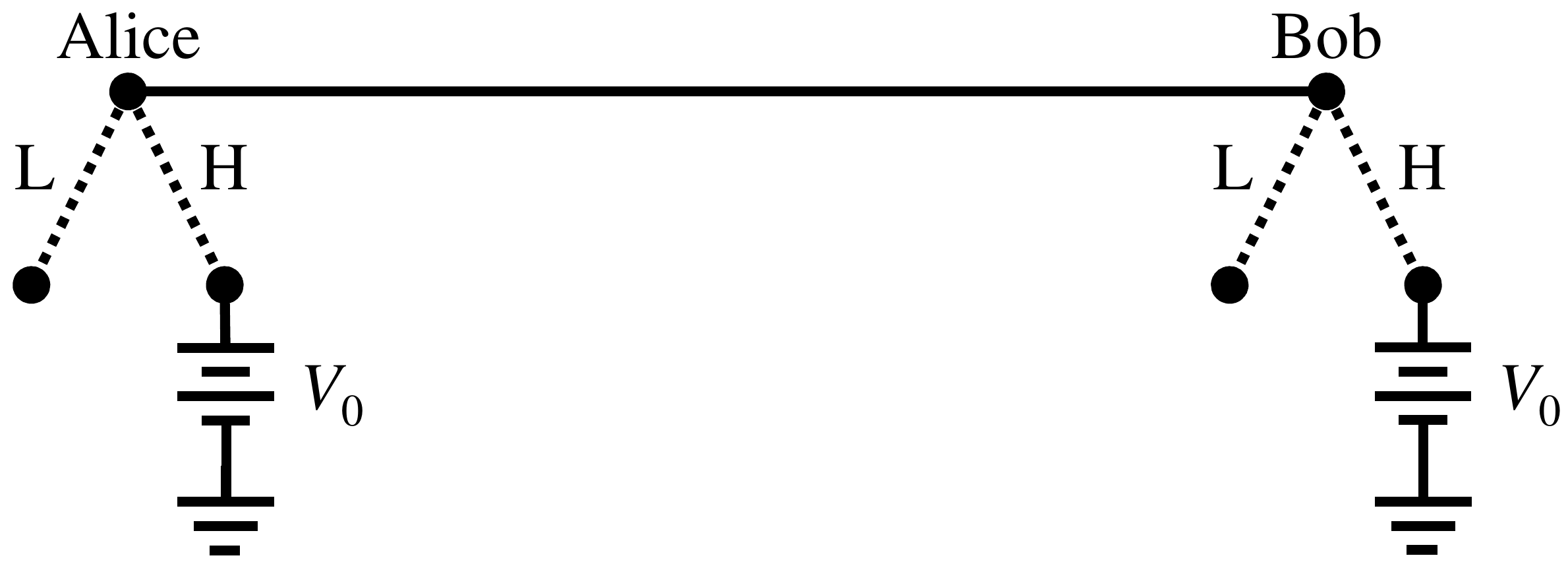}
\caption{\textbf{Noiseless KLJN protocol.} The wires and voltage sources are taken to be ideal, with zero thermal noise. For each clock cycle, Alice randomly chooses either ``L-first'' or ``H-first''. If the former, she connects the L side of her circuit for the first half of the clock interval, and the H side for the second half. (If the latter, she does the reverse.) Bob does likewise. If the choices are the same, then for half of the clock interval the voltage on the transmission line will be zero and that clock cycle is discarded.  If the choices are different, then the voltage is $V_0$ for the entire interval, regardless of who chose ``L-first''. (The transmission line can be momentarily re-grounded between half-clock cycles to ensure it has zero voltage when neither party connects a battery.) The choice is known to both Alice and Bob, which now forms their key bit, but is unconditionally secure from an eavesdropper according to the standards used to claim security of the original KLJN protocol in figure \ref{fig:originalkish}. Thermodynamics and noise do not play a role.}
 \label{fig:simplekish}
\end{figure}

First, we note that the security of the KLJN protocol is only claimed in the limit where electromagnetic modes of finite wavelength are neglected \cite{Kish2006a}. In other words, currents and voltages are assumed to be quasi-static. Kish has argued that this obviates any need to analyze waves moving along the transmission line \cite{Kish2006d} and he agrees intercepted waves would compromised security \cite{Kish2007}. We believe this no-wave limit is inappropriate and nonphysical for analyzing communication protocols (even as a mathematical idealization) because if propagating waves are excluded there is no way for information to get from Alice's side of the circuit to influence Bob's side, or vice versa. Unfortunately this no-wave limit has formed the basis of subsequent claims that the idealized protocol is unconditionally secure \cite{Kish2006d, Kish2011, *Kish2012} in response to criticism \cite{Scheuer2006}.

To illustrate our critique, we depict a simplified analog of the KLJN protocol in figure \ref{fig:simplekish} which is similarly secure against passive attack in the no-wave limit. Since the concepts of temperature and noise have been eliminated, they play no fundamental role in the protocol, undermining the recent claim of Kish et al.\ that the security of the KLJN protocol follows from the 2nd law of thermodynamics \cite{Kish2011, *Kish2012}. This was already suggested by the earlier observation in reference \cite{Kish2006a} that artificial noise generators were as good as true Johnson-Nyquist noise from resistors.

Kish has argued that many cryptographic schemes (such as quantum key distribution) were initially only analyzed as mathematical idealizations, and that the usefulness of the KLJN protocol can only be assessed with a detailed mathematical analysis of real-world inefficiencies \cite{Kish2006c, Kish2006d}. Although we are sympathetic to the general idea that physical robustness of information systems is very important and often non-obvious---witness fault-tolerant quantum computation \cite{Shor1996}---we emphasize that quantum key distribution \emph{has} been shown to be robust with imperfect components against very general attacks \cite{Mayers2001, *Gottesman2004, *Lo2012, *Braunstein2012} while exploiting an unphysical limit seems to be essential to the KLJN protocol. We can be more precise as follows.

Suppose that Eve makes a continuous measurement of the electromagnetic field on a thin cross-section of the transmission line (including the grounding wire) between times $t=0$ and $t=T$, denoted by the variable $Z$. This variable contains a complete history of the fields inside the cross-section during the interval of length $T$. Since this is a purely classical analysis, the measurement can be ideal. We can either (1) use the exact (or `microscopic') version of Maxwell's equations and imagine that this cross-sectional plane passes between all atoms in the transmission line or (2) use the macroscopic version of Maxwell's equations and take the transmission line to be a continuous medium. Either way, Eve can decompose the fields into orthogonal components $Z_\mathrm{A}$ and $Z_\mathrm{B}$, describing the waves heading toward Alice and Bob, respectively.

Let $Y$ be a variable that describes everything on Bob's side of Eve's location during the same time interval, including waves traveling toward him, away from him, and all of his equipment and memory. Let $X$ be the same for Alice. $X$, $Y$, and $Z = (Z_\mathrm{A}, Z_\mathrm{B})$ are random variables characterizing the distribution of possible histories when the protocol is run. (This captures, for example, the outcome of any coin flips Alice and Bob perform, as well as possible noise in the components.) The initial data on the $t=0$ surfaces of $X$ and $Y$ must be independent variables---otherwise, Alice and Bob could simply use them to construct a secret key---but $X$, $Y$, $Z_\mathrm{A}$, and $Z_\mathrm{B}$ will be generally all be correlated.

The deterministic and locally causal structure of Maxwell's equations ensures that $Y$ can only be influenced by $X$ via the intermediary $Z_\mathrm{A}$. This can be expressed mathematically as
\begin{align}
\label{eq:condent}
H(X \cond Z_\mathrm{A})=H(X \cond Z)=H(X \cond Z,Y).
\end{align}

The conditional information $H(S \cond T)$ denotes the uncertainty in $S$ remaining when $T$ is known. Equation \eqref{eq:condent} just says that, once $Z_\mathrm{A}$ is known, nothing further about $X$ is learned from measuring $Z_\mathrm{B}$ or $Y$. This can equivalently be stated as a restriction on the joint probability density:
\begin{align}
\label{eq:probs}
p(x,y,z) = \int \! \mathrm{d}z \, p(z) p(x \cond z_\mathrm{A}) p(y \cond z_\mathrm{B}),
\end{align}
where $z = (z_\mathrm{A},z_\mathrm{B})$ and where $p(z)$, $p(x \cond z_\mathrm{A})$, and $p(y \cond z_\mathrm{B})$ are defined in the usual way from $p(x,y,z)$. Intuitively, $X$ only ``knows'' about $Z_\mathrm{A}$ and so is completely specified by $p(x \cond z_\mathrm{A})$.

\begin{figure} [t]
    \centering
  \newcommand{\bwidthfactor}{0.65}
    \includegraphics[width=\bwidthfactor\columnwidth]{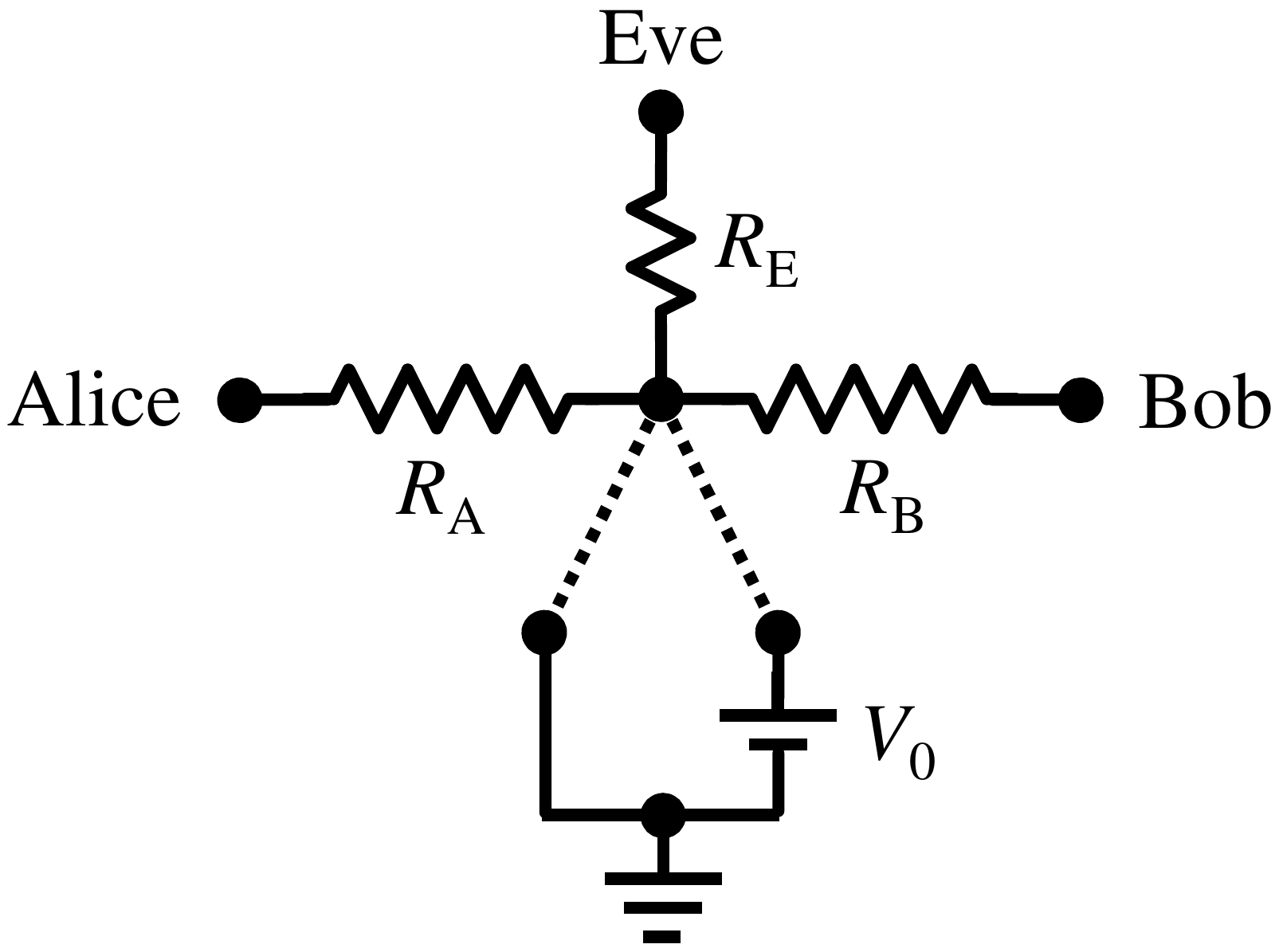}
\caption{\textbf{Key distillation from Johnson-Nyquist noise.} Consider a static lumped circuit with resistors at finite temperature connected as in the figure.  The switch for the central node is flipped randomly between ground and a (tiny) voltage $V_0$.  Alice, Bob, and Eve see this through noisy thermal resistors $R_\mathrm{A}$, $R_\mathrm{B}$, and $R_\mathrm{E}$, each of which applies an independent Gaussian offset to the voltage at the central node. Even if Eve's noise is less than Alice's and Bob's (e.g. if $R_\mathrm{E} < R_\mathrm{A}$, $R_\mathrm{E} < R_\mathrm{B}$, with everyone at the same temperature), Alice and Bob can use advantage distillation \cite{Maurer1993} to distill a secret key via 2-way public communication provided $R_E$ is strictly positive.  However if Eve's measurement of the fluctuating voltage at the central node is noiseless ($R_E=0$), Alice's and Bob's conditional mutual information, conditioned on Eve's observation, is zero, and no secret key can be distilled.}
  \label{fig:maurer}
\end{figure}

It follows that $X \to Z \to Y$ forms a Markov chain and that the conditional mutual information between $X$ and $Y$, conditional on $Z$, vanishes:
\begin{align}
\label{eq:mutualinfo}
I(X \bet Y \cond Z)=H(X \cond Z)-H(X \cond Z,Y)=0.
\end{align}
Under this condition Alice and Bob, even with the help of public discussion, cannot establish a key that is secret from Eve \cite{Maurer1999}. That is, the distillable key rate is zero.

This conclusion depends on Eve's ability to accurately measure the time history of the electromagnetic field, not just its instantaneous spectrum. Thus, while the steady state mean square noise voltage in the original KLJN protocol (figure \ref{fig:originalkish}) does not allow Eve to distinguish between the LH and HL settings of Alice's and Bob's resistors \cite{Kish2006a}, she can distinguish them using either (a) transient waves created by the switching action before the steady state is established, or (b) time correlations in the steady-state distribution of traveling waves resulting from the fluctuations that give rise to Johnson-Nyquist noise. For example Bob's resistor affects the phase and amplitude correlations between a right-traveling wave at time $t$ and its left-traveling echo at time $t+\Delta$, where $\Delta$ is the transit time from Eve to Bob and back, with the echo vanishing only if the resistor is perfectly impedance matched to his end of the line.

Our noiseless protocol in figure \ref{fig:simplekish} with ideal components has no fluctuations and is not open to this passive steady-state attack, but of course it could be broken by observing the transients as the voltage was being raised on the transmission line from one side or the other. These transients are precisely the waves which carry the information between Alice and Bob. However, if Eve graciously promises not to observe the transients, she could still learn the key by an \emph{active} steady-state attack in which she would place a very high-resistance shunt between her node and ground, and monitor the direction of current flow into it.   Of course Alice and Bob could try to detect this weak leakage current also, and abort the protocol if they found it.  The result would be an unstable arms race, won by whichever side had the more sensitive ammeter, not the sort of robustness reasonably expected of a practical cryptosystem.

It has been claimed that low-pass filters added by Alice and Bob in an attempt to enforce the no-wave limit would prevent attacks on propagating waves in KLJN systems \cite{Kish2006d, Kish2011, *Kish2012}. This claim has not been quantified, and in any case  filters added by Alice and Bob do not avoid our general theorem (\ref{eq:mutualinfo}) about classical electromagnetic signaling; any frequency components traveling toward Bob which are removed by Alice's low-pass filter before entering the public section of the transmission line are unavailable to both Eve and Bob.

We emphasize that so long as Eve makes perfect measurements, equation \eqref{eq:mutualinfo} holds regardless of whether $X$, $Y$, and $Z$ are continuous or discrete, whether Alice and Bob perform deterministic or probabilistically programmed actions specified by the protocol, and whether the equipment and transmission lines have loss, noise, filters, or even memory. Therefore, the problems of finite resistance in the transmission line \cite{Scheuer2006} or temperature differences between Alice and Bob \cite{Hao2006} are moot. Attacks based on these imperfections become important only when the unphysical no-wave limit is accepted, which as we noted would also prevent key agreement by preventing any information from passing between Alice and Bob in the first place.

Of course, none of this shows that ideas contained in the KLJN protocol cannot find useful application. There are many classical cryptographic protocols whose security rests on some assumed limitation on the class of attacks available to the eavesdropper. For example, the last two decades have seen interest in the use of synced chaotic lasers for transmitting messages which cannot be decoded by an eavesdropper using simple measurement strategies.  An early example relevant to key generation from thermal noise is the scenario illustrated in figure \ref{fig:maurer}, where Alice, Eve, and Bob listen to the same random binary source through three independent Gaussian channels, yielding variables $A$, $B$, and $E$.  Maurer showed (reference \cite{Maurer1993} section V) that Alice and Bob can distill secret key at a positive rate even if their channels are more noisy than Eve's by using two-way public communication to collaboratively agree on a subset of their data on which their noise is less than Eve's, even though it is greater than Eve's on the raw data as a whole.  The key rate obtainable from this ``advantage distillation'' technique depends on the noisiness of Eve's measurements, being upper bounded by the conditional mutual information $I(A;B|E)$, which vanishes in the limit $R_E\rightarrow 0$.   Advantage distillation has been further generalized and developed by Maurer and Wolf \cite{Maurer1999}.  For quantitative studies of key rates obtainable with discrete and Gaussian channels see \cite{Gander1994} and \cite{Naito2009}.

In conclusion, we have shown that protocols of the KLJN type do not use thermodynamics in an essential way, and that they can be completely defeated by an adversary able to perfectly monitor the time-dependent fields in the region between Alice and Bob, as classical electromagnetic theory would allow. On the other hand, if in practice Eve's monitoring can be guaranteed to be finitely noisy, such protocols may yield secure key, at a rate not exceeding the conditional mutual information, using Maurer's technique of advantage distillation.


We thank John Smolin and Graeme Smith for extremely helpful discussions. This work was supported in part by the John Templeton Foundation, grant number 21484.

\bibliographystyle{apsrev4-1}
\bibliography{riedelbib}

\end{document}